\documentclass{svjour3}                     
\smartqed  
\usepackage[normalem]{ulem}
\usepackage{graphicx}
\usepackage{color}
\usepackage{amssymb}
\usepackage{amsmath}
 \journalname{Scientometrics}

\definecolor{Gray}{gray}{0.7}

\begin{document}

\title{
Comparison of {a} citation-based {indicator} and peer review for
absolute and specific measures of research-group excellence
%
%
%
}


\author{O.~Mryglod \and R.~Kenna \and Yu.~Holovatch \and B.~Berche }

\institute{O. Mryglod \at
              Institute for Condensed Matter Physics of the National Academy of Sciences of Ukraine,
              1 Svientsitskii Str., 79011 Lviv, Ukraine\\
              \email{olesya@icmp.lviv.ua}
           \and
            R. Kenna \at
              Applied Mathematics Research Centre, Coventry University,
              Coventry, CV1 5FB, England
           \and
            Yu. Holovatch \at
              Institute for Condensed Matter Physics of the National Academy of Sciences of Ukraine,
              1 Svientsitskii Str., 79011 Lviv, Ukraine
           \and
            B. Berche \at
               Universit\'e de Lorraine, Campus de Nancy, B.P. 70239, 54506 Vand\oe uvre l\`es Nancy Cedex, France
}

\date{Received: date / Accepted: date}

\maketitle

\begin{abstract}
Many different measures are used to assess academic research excellence and these are subject to ongoing discussion and debate within the scientometric, university-management and policy-making communities internationally.
One topic of continued importance is the extent to which  citation-based indicators compare with peer-review-based evaluation.
Here we analyse the correlations between values of a particular citation-based impact indicator and peer-review scores in several academic disciplines, from  natural to social sciences and humanities.
We perform the comparison for research {\emph{groups}} rather than for individuals.
We make comparisons on two levels.
At an  absolute level, we compare total impact and overall {\emph{strength}} of the group as a whole.
At a specific level, we compare academic impact and {\emph{quality}}, normalised by the size of the group.
We find very high correlations at the former level for some disciplines  and poor  correlations at the latter level  for all disciplines.
This means that, although the citation-based scores could help to describe research-group strength, in particular for the so-called hard sciences, they should not be used as a proxy for ranking or comparison of research groups.
Moreover, the correlation between peer-evaluated and citation-based scores is weaker for soft sciences.
\end{abstract}

\section*{Introduction}

Although it is not without critics, peer-review is mostly considered, amongst the broad academic community, to be the most reliable approach to assess the quality of academic research \cite{2005_Raan,Derrick}.
However because it is expensive, time-consuming and dependent on different circumstances (the so-called Hawthorne effect, see Ref.~\cite{2012_Bornmann}), it is tempting to seek {other approaches} and citation-based indicators offer an obvious alternative \cite{nature,2003_Warner}.
{Numerous scientometric indicators based on the citations number as well as the general number of publication and other aspects were proposed during the past half of century (e.g., see \cite{Garfield_1955,Garfield_1973,Hirsch_2005,Egghe_2006}). The real challenge is to invent a simple but reliable way to assess the individual or collective scientific performance. The sophisticated normalization procedures and different approaches have been designed to overcome the well-known nuances of citations counts \cite{Vinkler2001,Vinkler2003,2005_Moed}. But this remains the problem of current importance till today.}
For over half a century, scientists and research managers have discussed the merits and drawbacks of each approach.
For practicing academics the accuracy and reliability of peer review broadly wins out
(see, e.g., Refs.~\cite{Derrick,Donovan,Bornmann} and references therein).
University managers, policy makers and the media, however, are attracted to the simplicity and economy of citation-based methodologies.
Each approach is beset by  ambiguities and problems and it is frequently argued that a combination may be needed to minimize the shortcomings of each.
To achieve this, the technical and methodological limitations need to be clear \cite{2005_Raan}.
Here we address the question of whether a set of automated, scientometric or bibliometric indicators is a suitable substitute for, or component of, peer-review {\emph{at the level of the research group or department.}}

The importance of evaluation of research quality at institutional level is exemplified by the growing number of reports produced by private companies and governmental bodies which document research performance of Higher Education Institutions  within nations and worldwide
(e.g., \cite{2010_Nature,2012_Melbourne_report,2012_Evidence}).
The {\emph{Research Assessment Exercise}} (RAE) and {\emph{Research Excellence Framework}} are examples of such processes nationally in the UK, and the Shanghai Academic Ranking is a famous example on an international scale~\cite{Shanhai}.
{The Shanghai Ranking, in particular, is widely known but heavily criticised by the scientometric community \cite{Florian,Billaut,Ioannidis}. }
Despite well-known weaknesses of different systems for ranking universities, these are  of increasing importance in many developed countries, which seek to organize national assessments of research. Many aspects of the  UK's RAE, in particular, have been imitated in other countries \cite{2010_2_Nature}.

In a recent paper \cite{2012_Scientometrics} we compared a citation-based indicator developed by \emph{Thomson Reuters Research Analytics} (previously known as \emph{Evidence})  \cite{Evidence_web}
to the peer-review-based RAE which was conducted  in the UK in 2008.
\emph{Thomson Reuters} is one of the world's leading providers of scientometric information and performance measures for academic and research institutions, governments, not-for-profit organisations, funding agencies, and  others with a stake in research. 
E.g., {\emph{Thomson Reuters}}' (formerly the {\emph{Institute for Scientific Information}}) {\emph{Web of Knowledge}} is an important  platform for information on citations in the sciences, social sciences, arts, and humanities.
Using biology research institutions  as a test case, we examined the correlations between results from both approaches at an amalgamated, research-group or department level. We made the comparison at two levels which we termed  ``absolute'' and ``specific''. ``Absolute'' measures refer to the totality of group strength -- the research performance of the group as a whole. ``Specific'' quantities are normalised per head -- the average strength, taken per group member. In this sense,  ``absolute strength'' is the ``volume of quality''.
E.g., the {\emph{absolute}} citation count for a department in a given period is the total number of citations to the department's work, irrespective of how many researchers that department contains. The corresponding {\emph{specific}} citation  count is then the average number of citations per head  (see, for example, \cite{Vinkler2001,Vinkler2003}).

Thus, the estimates of research ``quality'' and research ``strength'' introduced in \cite{2010_Ralph,2011_Ralph} are specific and absolute notions, respectively.
We showed that the citation-based \emph{specific} measure $i$ provided by {\emph{Thomson Reuters Research Analytics}} is not a good proxy for the peer-review \emph{specific} measure $s$ provided by RAE, in that these two  measures are rather poorly correlated.
However, when scaled up to the actual size $N$ of a department %
{(here and below $N$ means the number of researchers in group)}%
, the {\emph{absolute}} citation impact ${\mathcal{I}}=iN$ is very strongly correlated with the overall strength ${\mathcal{S}}=sN$ as measured by peer review.
This is important because {funding in the UK} is determined on the base of strength ${\mathcal{S}}$ rather than quality $s$.

{Another important feature of our previous analyses was that they focused on the research quality and strength of {\emph{groups}} rather than individuals \cite{2012_Scientometrics,2010_Ralph,2011_Ralph}.
In particular, the notion of} two characteristic group sizes {or} ``critical masses'' {was} introduced in Refs.~\cite{2010_Ralph,2011_Ralph}.
According to this concept, {research performance is strongly dependent on group size up to a so-called upper critical mass $N_{\mathrm{c}}$. Groups  larger than} $N_{\mathrm{c}}$ have either a reduced dependency of quality on quantity or no such dependency.
{A} lower critical value $N_{\mathrm{k}}$ was {also introduced in Refs.\cite{2010_Ralph,2011_Ralph} and} interpreted as the minimum size a research department should achieve to be stable in the long term.
These two critical masses, the values of which are strongly dependent on the research discipline, allow research groups and departments to be categorised as being {\emph{small}} if they have size $N \leqslant N_k$, {\emph{medium}} if $N_k \leqslant N \leqslant N_c$ or {\emph{large}} if $N>N_c$.
E.g., for the biological sciences analysed in {the pilot study of} Ref.~\cite{2012_Scientometrics},
the estimates for critical masses are $N_k = 10.4$ and $N_c = 20.8$ \cite{2010_Ralph,2011_Ralph}.
(Fractions of staff are a feature of RAE in that Higher Education Institutes can include part-time researchers in their submissions and are counted as a proportion of full time equivalence \cite{RAE_web}.)
However, since small and medium research groups  have the same linear dependency of quality on quantity \cite{2010_Ralph} it is sensible to combine them in the correlation analysis.
The strongest correlations between citation- and peer-review based measures of institutional strength for the biological sciences was observed for the large groups.

The implication of our previous analysis, therefore, is that citations, if used in an informed manner, could possibly be used as a proxy for departmental or group {\emph{strength}} (and thus {feed into funding requirements}), provided that the departments are large.
For smaller departments, however, peer review remains essential to determine {\emph{strength}}.
Moreover, citation-based indicators should not be used {{\emph{in isolation}}} to estimate  research {\emph{quality}} {for large,  medium or for small groups.}

It is natural to ask to what extent these conclusions cover other disciplines.
Is there a difference between so-called hard and soft sciences or between the
natural and social sciences and humanities?
One might expect to observe differences due to different citation behaviour in different disciplines  \cite{2005_Moed,2012_Stauffer} and due to technical restrictions such as a smaller coverage by the {\emph{Web of Knowledge}}.
E.g., in the humanities, dissemination of original research through books is more common  than in the natural sciences, and these are usually ignored in citation counting. These are the questions we address in this paper.
We present quantitative results from comparisons of peer review and citation-based
indicators for several disciplines ranging from hard sciences to humanities. In particular, we consider chemistry; physics; mechanical, aeronautical and manufacturing engineering; geography and environmental studies; sociology and history.

 Again, as in Ref.~\cite{2012_Scientometrics}, we used data from {\emph{Thomson Reuters Research Analytics}} and the UK's {2008 version of the \emph{Research Assessment Exercise} (called RAE~2008).}
As in the pilot study for biology, here we provide evidence that
correlations between {\emph{specific}} citation indicators and peer-measured group qualities for all the disciplines are very weak, even in the case of ranked values. However, when scaled up to the actual size of the department $N$, the absolute citation impact is strongly correlated with the overall group strength as measured by peer review. The correlation is very strong (above 95\%) for the hard sciences,
less strong for geography and engineering, and weakest for social sciences
(below 90\%).
Although the correlations of measures are statistically strong for all the disciplines examined, since national assessment is linked to  funding distribution, even small differences can involve large financial impact. Thus, the threshold of reliability of results should be very high.
This means that our previous conclusions {\cite{2012_Scientometrics}} indeed extend to the hard sciences, physics and chemistry. But they do not extend to beyond the natural sciences.
The social sciences and humanities, in particular, require peer-evaluated measurements of both quality and strength.


\section{Peer review and the {N}ormalised {C}itation {I}mpact for research institutes}

\subsection{The Research Assessment Exercise and the Research Excellence Framework}

Quality related funding forms one element of the UK's dual research-support system.
Until now, this has been based on the {RAE} \cite{RAE_web} and the annual distribution of quality-related funding is over  2 billion euro.
In the future it will be based on the \emph{Research Excellence Framework}  \cite{REF_web}.
The evaluation of the quality of academic research output forms the major component of each of these schemes.
Using published criteria, RAE~2008 assessed submissions in each of 67 different subject areas (units of assessment) and awarded a profile for each of them.
{All submissions are related to the assessment period which is from 1 January 2001 to 31 July 2007 \cite{RAE_web}.
Submissions included four outputs (publications) per staff member.
E.g., in physics 1686 scientists submitted to the RAE.
This involves 6744 papers. (The actual number may be somewhat less than this because co-authored papers should be attributed proportionally to each contribution.) There was an average of 40 authors per submission, which translates into 160 papers per group.
RAE experts seek to quantify the proportion of a department's or research centre's submitted work which falls into each of five quality bands.}
The highest band is denoted as 4* and represents world-leading research.
Remaining bands are graded  through  3*, 2* and 1* to the lowest quality level which is called ``Unclassified''~\cite{RAE_statement_panelE}.
The RAE quality profile assigned to a given research group is represented by a set of values $p_{n*}$, which  represent the percentage of a team's research which was rated $n*$. For example, the profile $p_{4*}=25,p_{3*}=20,p_{2*}=35,p_{1*}=15,p_{U}=5$ would indicate that 25\% of a groups research is of world-leading quality; 20\% is of 3* (internationally excellent); 35\% is of 2* quality (recognised internationally)
and 15\% is 1* (recognised nationally).

Governmental funding post RAE is determined by a formula which combines the  quality scores in a weighted manner.
While the formula is subject to regional and temporal variation (the latter often due to the influence of lobby groups) the one introduced by the {\emph{Higher Education
Funding Council for England}}  immediately following RAE 2008 rated
4* and 3* research as being seven and three times the value of 2* work, while lower quality research was unrewarded \cite{HEFCE_web}.
%
In Ref.~\cite{2012_Scientometrics}, we denoted the strength of a given research group by ${\mathcal{S}}$. This is defined as  the volume of quality,
\begin{equation}
 {\mathcal{S}} = sN,
\end{equation}
where $N$ is the size of the group of quality $s$.
The  amount of quality-related funding distributed by the {\emph{Higher Education
Funding Council for England}} to a given university after RAE is a function of its strength ${\mathcal{S}}$.
While strength determines future funding,
it is, of course, not sensible to rank groups or universities according to their
${\mathcal{S}} $ values because different group shave different sizes.
However many media and managers readily rank according to the quality measures $s$
(although this also neglects very strong size effects {as pointed out in Refs.~\cite{2010_Ralph,2011_Ralph}}).

At RAE, the  overall quality profile is constructed by summing sub-profiles for three separate elements (quality of ``outputs'', quality of ``environment'' and quality of ``esteem''), of which outputs play the strongest role.
In the future, the {{Research Excellence Framework}} will replace the RAE for peer-review, institutional research assessment  \cite{REF_web}.
The main difference is that overall quality profile will consist of ``outputs'', ``impact'' and ``environment'' instead of ``outputs'', ``esteem'' and ``environment''.
Here ``impact'' refers to non-academic impact (thus, not e.g., citations).
This new element is one of the major innovations of {{Research Excellence Framework}}.
But obviously, the very question about applicability of scientific results since long ago has been considered as one
of the aspects of scientific productivity.
Nevertheless, the ``outputs'' sub-profile remains the most important component of research assessment within {{Research Excellence Framework}} providing the 65\% of Overall score.
(The remaining 35\% is distributed between ``impact'' (20\%) and ``environment'' (15\%)  \cite{REF_web}.)
To summarise, peer-review measures of research outputs will continue to dominate the UK's   assessment of institutional research quality and strength in the years to come, and will be the main factor upon which billions of euros worth of funding will be allocated.

Although they may be influenced by non-academic impact and environment (e.g., visibility), citation counts refer only to outputs.
Therefore it is sensible to compare  citation-based measures with the ``outputs'' category of RAE.
These are readily available on the official RAE 2008 web-page \cite{RAE_web} and we will henceforth confine our attention to these measures.
To maintain consistency of notation with respect to Ref.~\cite{2012_Scientometrics},
we denote by $s_1$ the peer-review measure of quality coming from the ``outputs'' category of RAE~2008. The corresponding absolute measure is denoted by ${\mathcal{S}}_1=s_1 N$.

\subsection{Thomson Reuters Research Analytics citation indicator}

As described in Ref.~\cite{2012_Scientometrics}, our citation-based measure of choice is that provided by {\emph{Thomson Reuters Research Analytics}}.
This company offers a service analysing research performance tailored to individual client requirements \cite{Evidence_web}.
They have developed the so-called {\emph{Normalised Citation Impact}} (NCI) $i$ as a coefficient of departmental performance in a given discipline.

{\emph{Thomson Reuters Research Analytics}}  calculate the  NCI using data from {\emph{Web of Knowledge}} databases \cite{Evidence_2010,Evidence_2011}.
Similarly to {\emph{Relative Citation Rate}} (RCR) (i.e., \cite{Schubert1996}), {the} NCI is calculated by comparing to a mean or expected citation rate.
It is  a {\emph{specific}} measure of  {academic citation} impact because it is averaged over the entire research group.
A  non-trivial advantage of the NCI is that it takes account of different citation patterns between different academic disciplines.
To achieve  this, the total citation count for each paper is
first normalised to an average number of citations per paper for the year of
publication and either the field or journal in which the paper was
published. This is called ``rebasing''  the citation count \cite{Evidence_2011}.
To compare sensibly with the UK's peer-review mechanism, only the four papers per individual which were submitted to RAE 2008 were taken into account by {\emph{Thomson Reuters Research Analytics}} in order to determine the average NCI for research groups \cite{Evidence_2011} {(citation data till the end of 2009 were analysed, see \cite{Evidence_2011}, Appendix A).}

Thus, the NCI may be considered as a citation-based {\emph{specific}} measure of
the academic impact of a department in a given field and we denote
it by $i$. The corresponding {\emph{absolute}} measure of impact
(the total volume of academic impact of the department or group) is denoted by
${\mathcal{I}}$. The relationship between the two is
\begin{equation}
{\mathcal{I}}=iN.
\end{equation}

\subsection{Comparisons to be made}
The objective of the remainder of this paper is to compare the
peer-review and citation-based indicators for different disciplines.
The {\emph{specific}} indicators to compare are quality and citation impact $s_1$ and $i$ as measures of the average strength and impact of the group or department {\emph{per individual}} contained within it.
We also compare the {\emph{absolute}} indicators ${\mathcal{S}_1}$ and ${\mathcal{I}}$ as measures of the overall strength and total impact of the group as a whole.

\section{Weak correlation between specific measures of quality and impact}

\begin{figure}[!h]
\centerline{\includegraphics[width=0.42\textwidth]{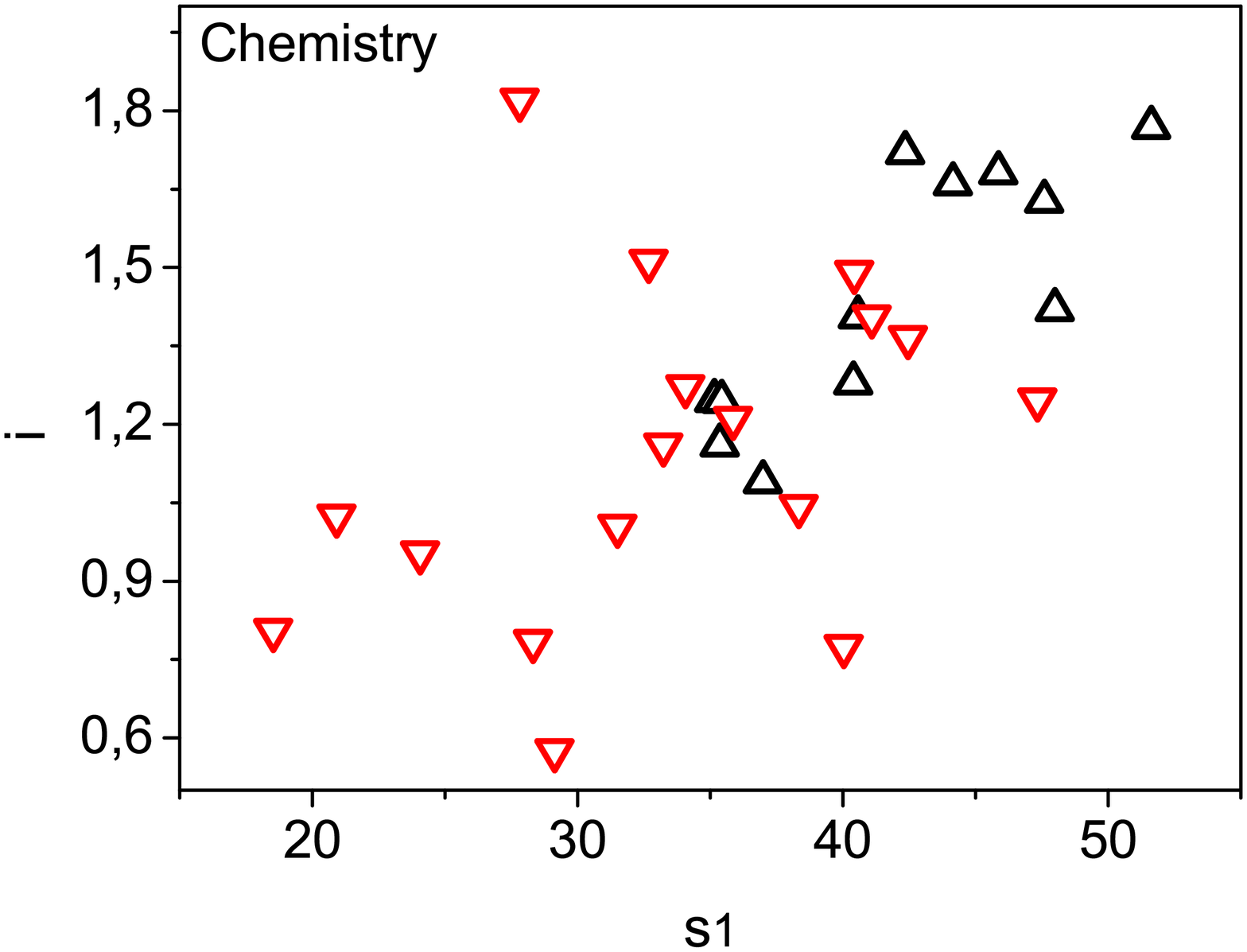}
\hspace{-11mm}
\includegraphics[width=0.42\textwidth]{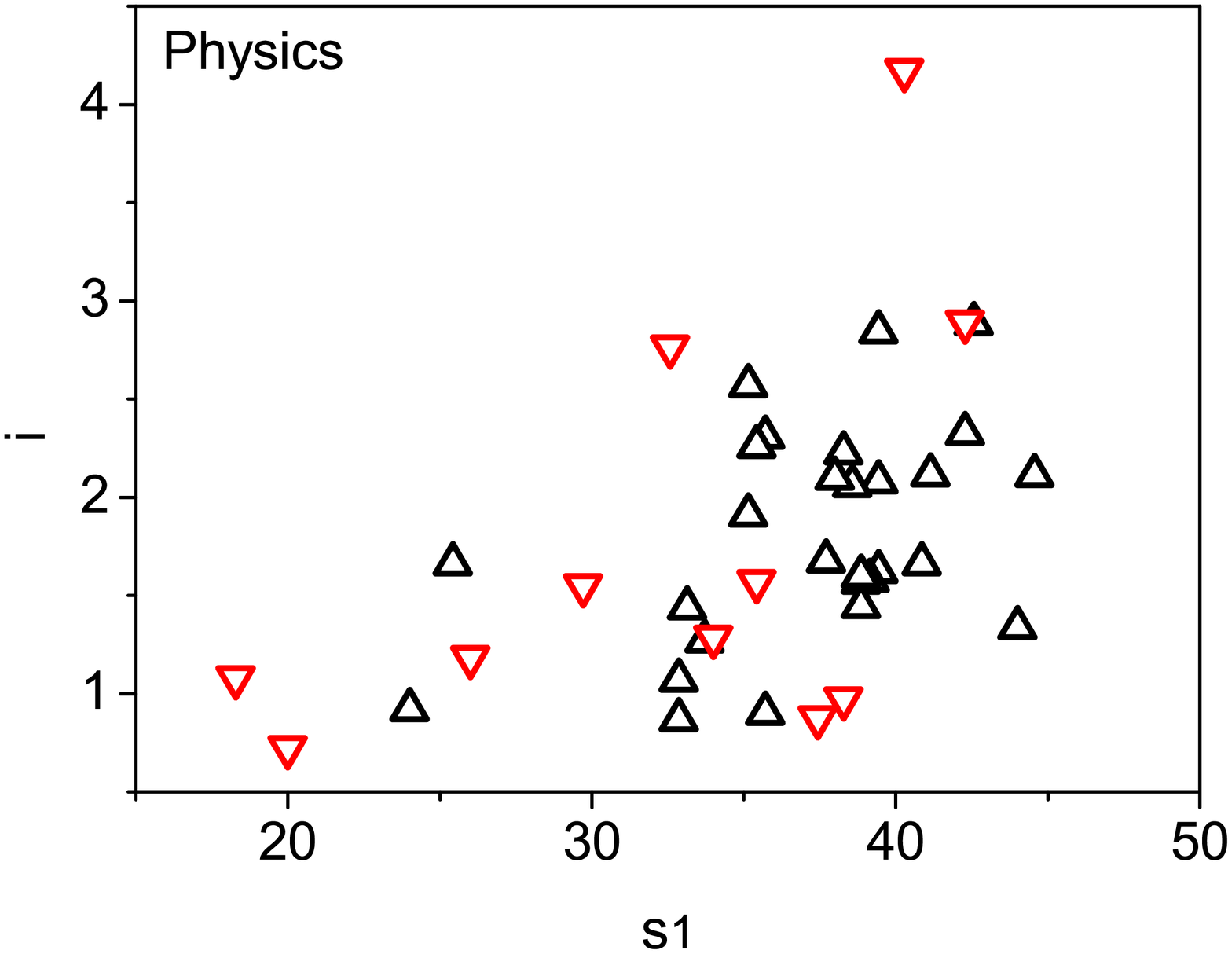}
\hspace{-11mm}
\includegraphics[width=0.42\textwidth]{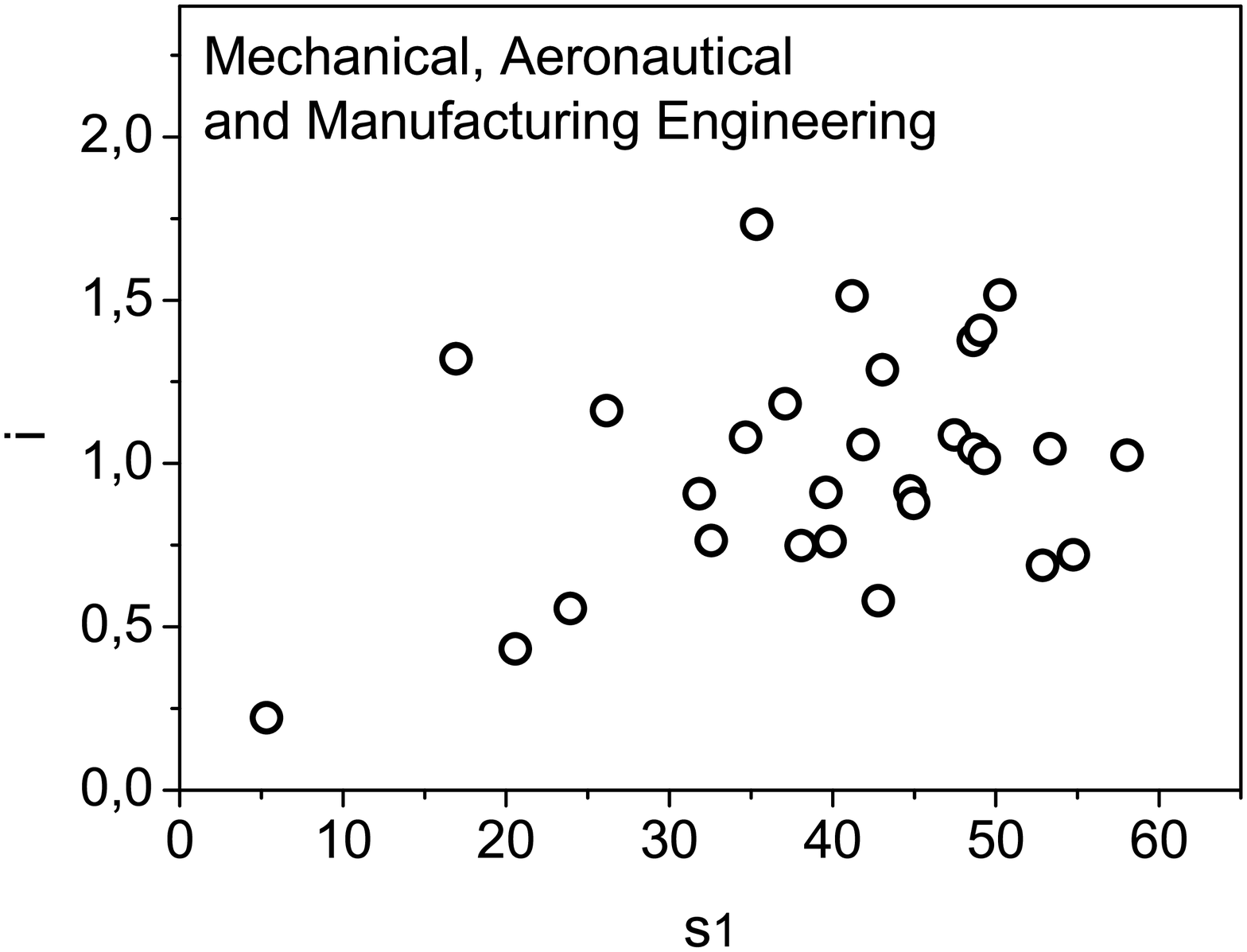}}
\centerline{(a)\hspace{4cm} (b)\hspace{4cm} (c)}
\centerline{\includegraphics[width=0.42\textwidth]{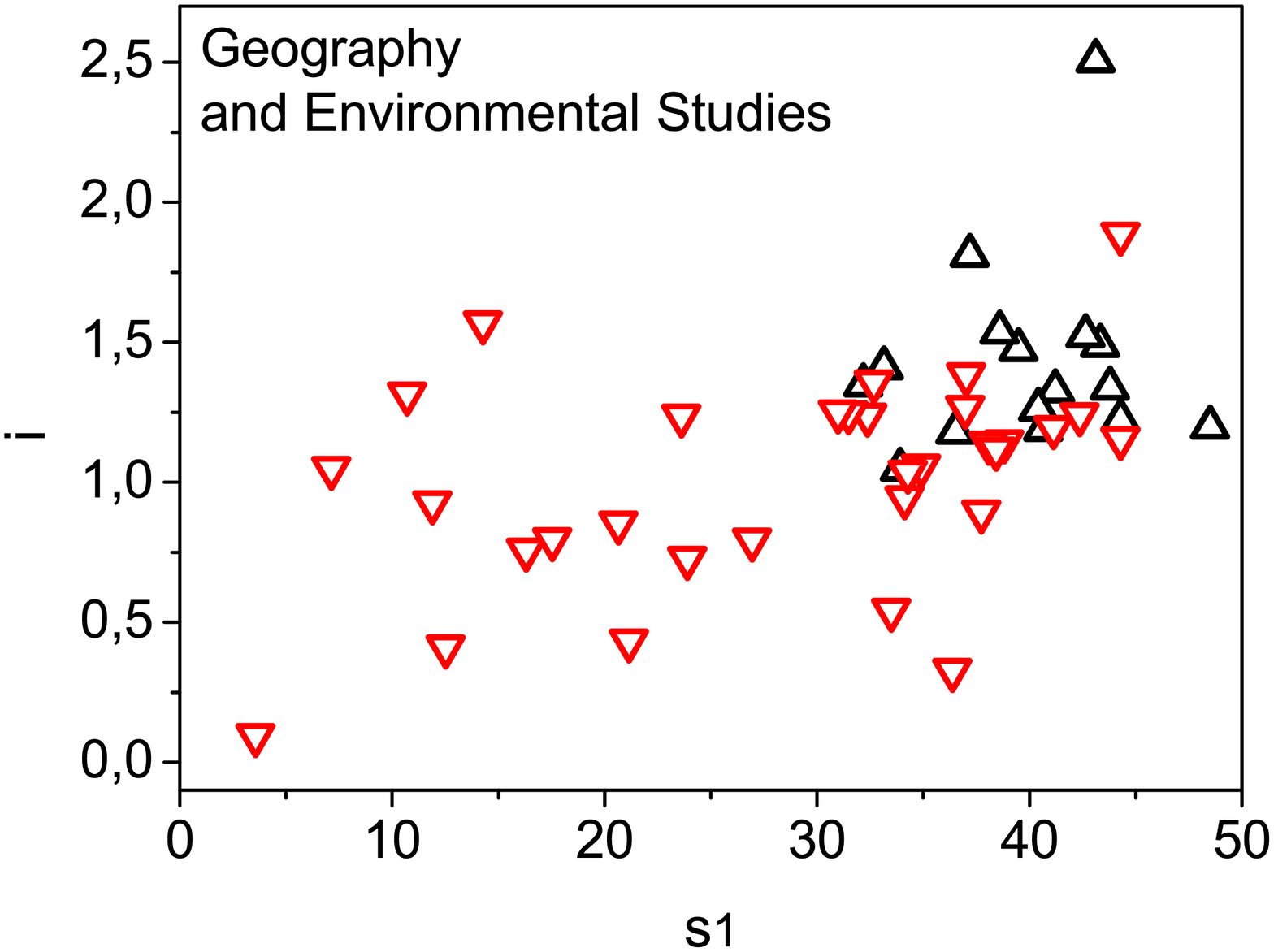}
\hspace{-11mm}
\includegraphics[width=0.42\textwidth]{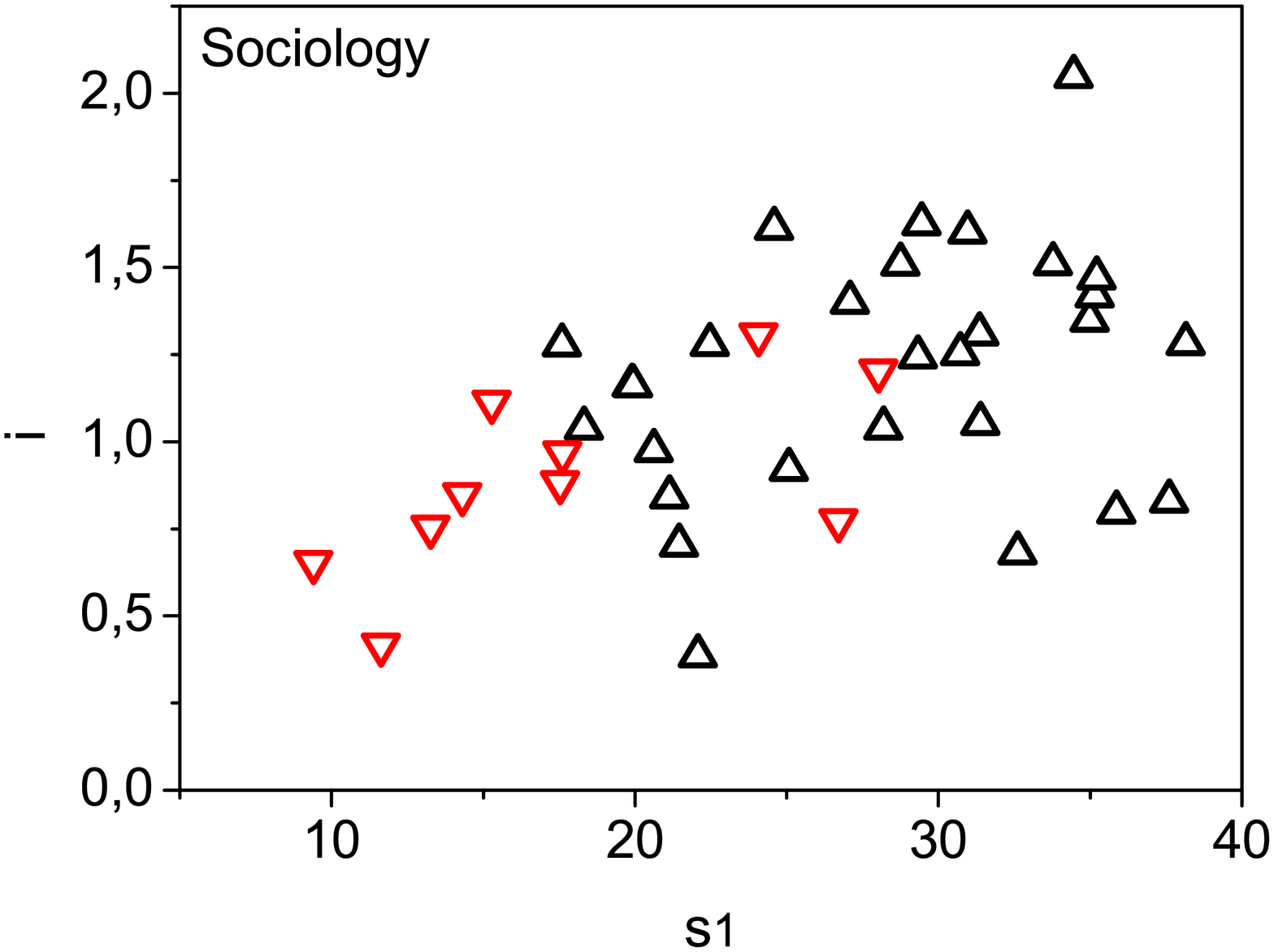}
\hspace{-11mm}
\includegraphics[width=0.42\textwidth]{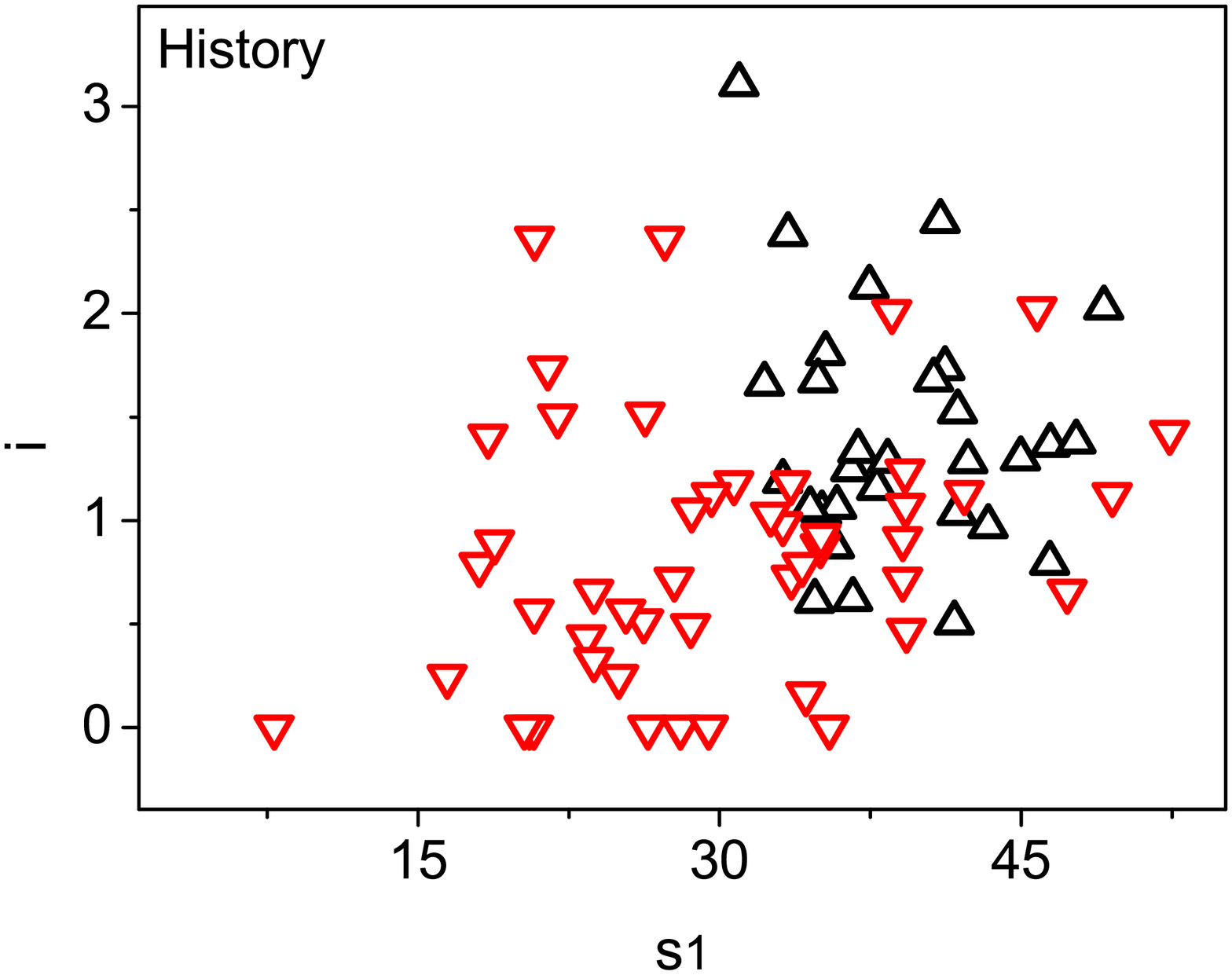}}
\centerline{(d)\hspace{4cm} (e)\hspace{4cm} (f)}
\caption{Correlations between average quality of research groups according to RAE 2008 $s_1$ and average excellence of research groups according to Normalised Citation Impact $i$ for: (a) chemistry, (b) physics, (c) mechanical, aeronautical and manufacturing engineering, (d) geography and environmental studies, (e) sociology and (f) history. Different symbols represent large (black $\triangle$) and  medium/small (red $\bigtriangledown$) groups. For engineering (c) information about group sizes is unavailable.} \label{fig1_RAE_vs_Evidence}
\end{figure}

A 100\% linear correlation between  $i$ and $s_1$ would indicate that
the citation-based  indicator (NCI) is a  perfect proxy for RAE peer-review quality scores.
The actual correlations for different disciplines are depicted in
Fig.~\ref{fig1_RAE_vs_Evidence} and are far from perfect.
For the majority of disciplines one can observe some positive but weak correlation.  This is quantified by a relatively small values of the Pearson
coefficient $r$, the values of which are listed in Table~\ref{tab_coefficients_1}.
The conclusion is clear -- the NCI  indicators should not be used in place of
peer-review measures of research-output quality.

As stated, normalized scores (be they RAE quality measurements or NCI citation-based indicators) are frequently  used for ranking research groups.
For this reason we also check the correlation between {\emph{ranks}}.
The ranks are constructed by listing ratings of research groups ascending order of their corresponding scores. Then each department is assigned an ascending numerical rank (the average rank in the case of the equal scores). The linear correlation strength between ranked variables is expressed by Spearman coefficient $\rho$ and these are also listed  in Table~\ref{tab_coefficients_1}.
\begin{table}[!ht]
\caption{The approximate values of linear correlation coefficients between specific values $s_1$ and $i$ calculated for several different disciplines. Statistically significant values are highlighted in boldface  (with significance level $\alpha=0.05$).}
\begin{center}
\begin{tabular}{|l|l|l|l|l|}
\hline
Description of the  & \multicolumn{3}{|c|}{Pearson coefficient $r$}& Spearman\\
\cline{2-4}
data sets &all &large  &medium& coefficient\\
&groups&  groups&/small & of ranked\\
&&   &  groups& values $\rho$  \\
\hline
\hline
biology (see Ref. \cite{2012_Scientometrics})&
{\mathversion{bold}$0.60$}& 
{\mathversion{bold}$0.57$}& 
$0.35$&
{\mathversion{bold}$0.53$}\\
(44 groups: 32 large, &&&&\\
7 medium, 5 small) &&&&\\
\hline
chemistry&
{\mathversion{bold}$0.60$}& 
{\mathversion{bold}$0.82$}& 
$0.34$&
{\mathversion{bold}$0.62$}\\
(29 groups: 12 large, &&&&\\
14 medium, 3 small)& &&&\\
\hline
physics&
{\mathversion{bold}$0.48$}& 
{\mathversion{bold}$0.45$}& 
$0.54$&
{\mathversion{bold}$0.53$}\\
(41 groups: 28 large, &&&&\\
9 medium, 4 small)&&&&\\
\hline
mechanical, aeronautical &
$0.34$& 
--& 
--&
$0.18$\\
and manufacturing engineering&&& &\\
(30 groups)& &&&\\
\hline
geography and environmental &
{\mathversion{bold}$0.51$}& 
$0.13$& 
{\mathversion{bold}$0.42$}&
{\mathversion{bold}$0.47$}\\
studies & &&&\\
(41 groups: 28 large, &&&&\\
9 medium, 4 small)& &&&\\
\hline
sociology &
{\mathversion{bold}$0.49$}& 
$0.29$& 
{\mathversion{bold}$0.64$}&
{\mathversion{bold}$0.47$}\\
(39 groups: 29 large, &&&&\\
8 medium, 2 small) &&&&\\
\hline
history &
{\mathversion{bold}$0.34$}& 
$<0$& 
$0.27$&
{\mathversion{bold}$0.38$}\\
(79 groups: 30 large, &&&&\\
24 medium, 25 small) &&&&\\
 \hline
\end{tabular}
\label{tab_coefficients_1}
\end{center}
\end{table}

Contrary to some earlier results which claimed high levels of correlation between the ranked RAE scores and citation counts (for example, $\rho\approx 0.80$ for music \cite{2008_Oppenheim} and $\rho\approx 0.81$ for archaeology \cite{2003_Norris}), our values of Spearman coefficient are low, varying from 0.18 to 0.62.
This is perhaps unexpected, since the normalised citation impact $i$ is a more sophisticated citation-based measure of academic impact compared to simple citation counting which was used in earlier analyses.

As stated earlier, it was established in Refs.~\cite{2010_Ralph,2011_Ralph} that the dependency of research quality on quantity of researchers differs depending on whether or not research  groups exceed the upper critical mass $N_c$. For this reason, we also investigate these categories (large and medium/large groups) separately.
The correlation coefficients between $s_1$- and $i$-values for large and for  medium/small groups are also listed in Table~\ref{tab_coefficients_1}.
As one can see, the proportions of groups with $N>N_c$ and $N < N_c$  differ  for different disciplines: whereas the sociological groups are mainly large, there is a high proportion of small/medium groups in the field of geography and environmental sciences.
However, this division does not help and the correlation coefficients for the two {\emph{specific}} measures of group-research performance are poor.
(Further subdivision of the $N<N_c$ category into separate sets of small and medium sized groups does not ameliorate the situation.)
We conclude that the NCI is a poor proxy for peer review measures of research quality in all subject areas analysed.


\section{Strong correlation between absolute measures of strength and impact}

A conspicuous feature of the above analysis is that all research groups are treated as contributing the same weight to the analysis.
For example, the RAE-measured quality scores for the history-research groups at the Open University and the University of Glamorgan are almost equal: $s_1 \approx 34$.
But, with 20.6 staff, the former is more than 3 times bigger than the latter which has only  6 researchers.
This means that researchers in smaller groups contribute more weight to the analysis, and statistical inaccuracies in their scores are unduly amplified.
This problem is remedied by multiplying the average quality of groups by their size, a process which also  renders the specific measures absolute: quality becomes strength and the NCI is also scaled up to the volume of the group or department.
\begin{figure}[ht]
\centerline{\includegraphics[width=0.42\textwidth]{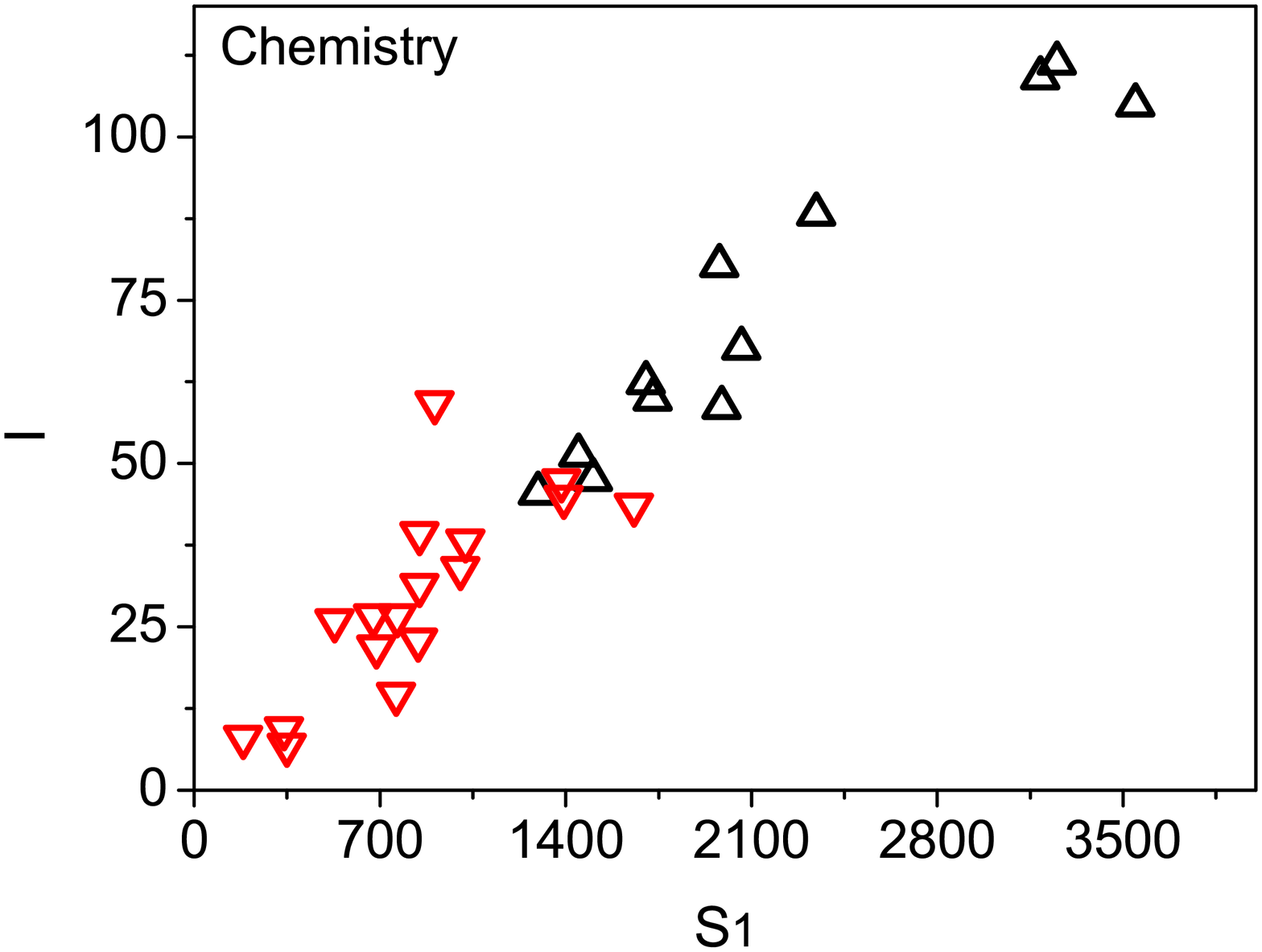}
\hspace{-11mm}
\includegraphics[width=0.42\textwidth]{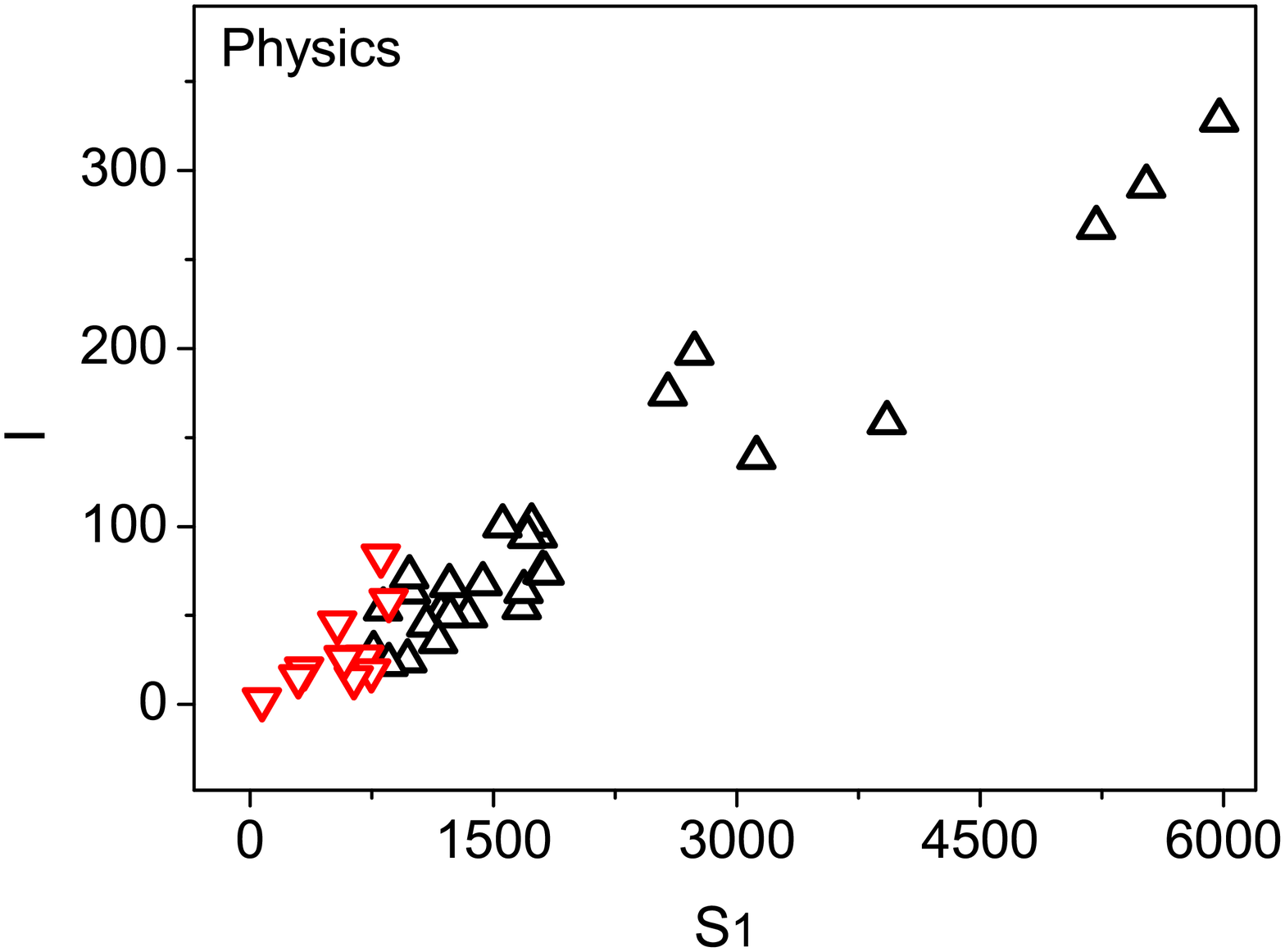}
\hspace{-11mm}
\includegraphics[width=0.42\textwidth]{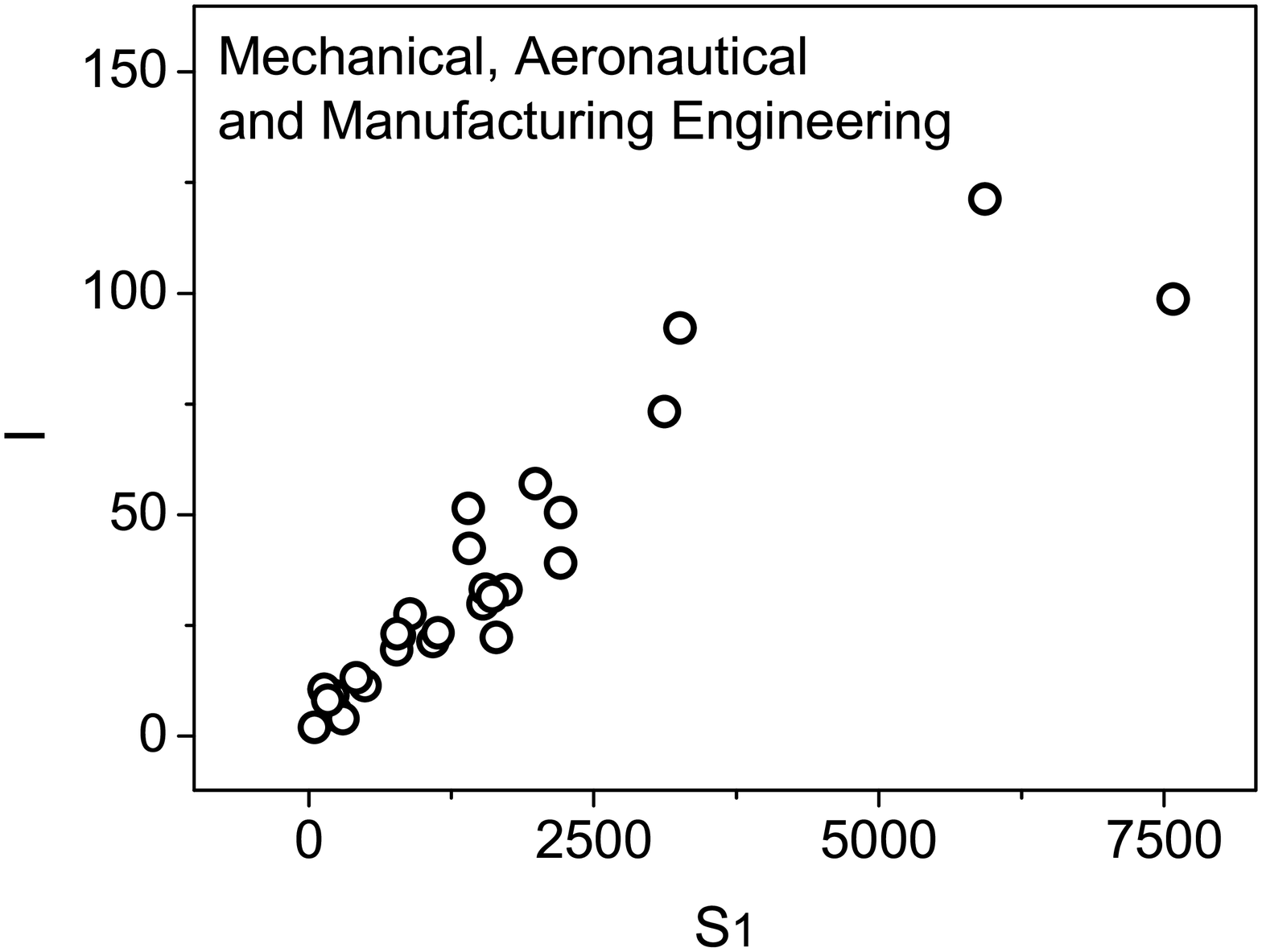}}
\centerline{(a)\hspace{4cm} (b)\hspace{4cm} (c)}
\centerline{\includegraphics[width=0.42\textwidth]{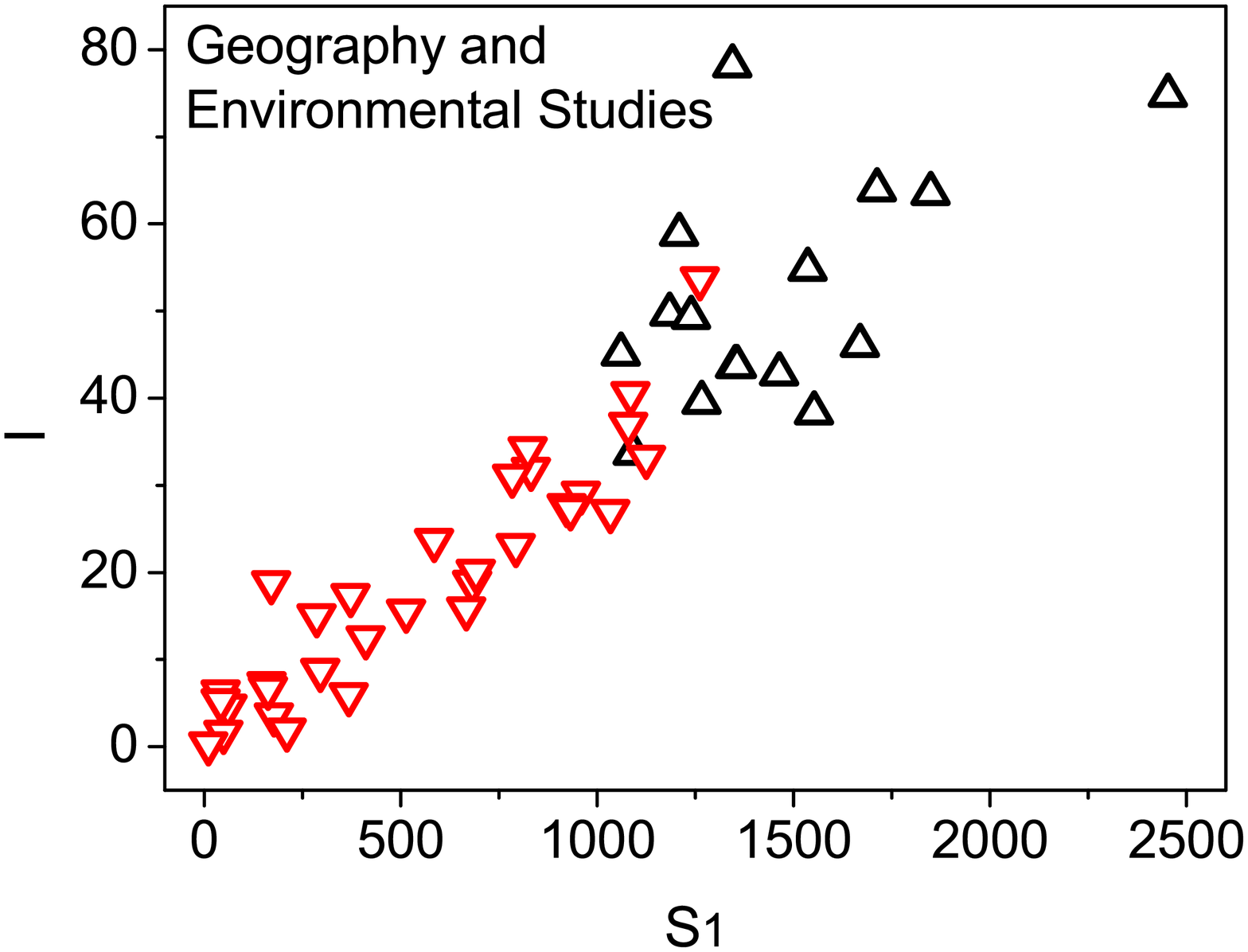}
\hspace{-11mm}
\includegraphics[width=0.42\textwidth]{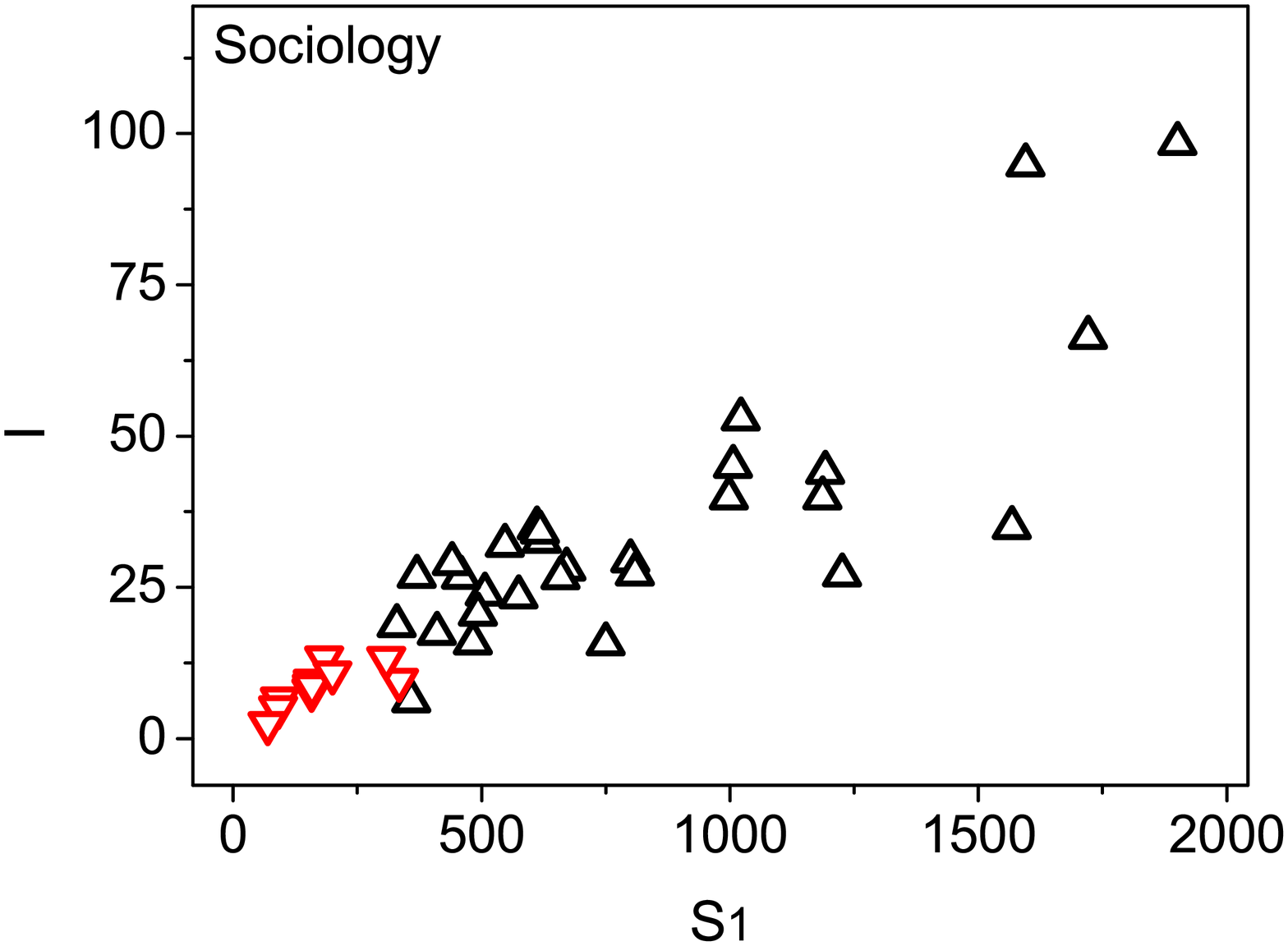}
\hspace{-11mm}
\includegraphics[width=0.42\textwidth]{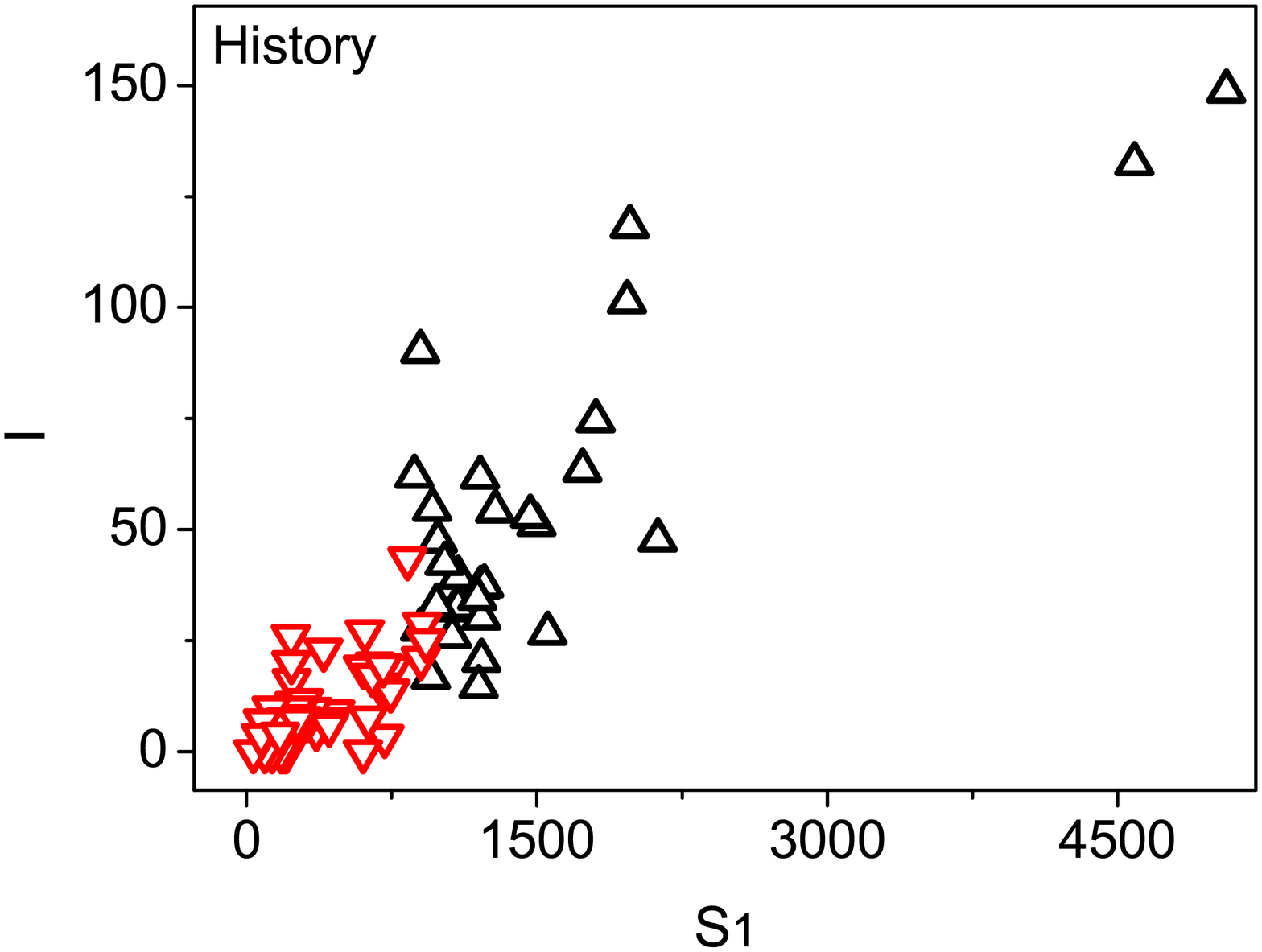}}
\centerline{(d)\hspace{4cm} (e)\hspace{4cm} (f)}
\caption{Correlation between ${\mathcal{S}_1}$ (strength of research groups according to RAE 2008) and ${\mathcal{I}}$ (absolute  citation impact) for: (a) chemistry, (b) physics, (c) mechanical, aeronautical and manufacturing engineering, (d) geography and environmental studies, (e) sociology and (f) history. The symbols are as in Fig.~\ref{fig1_RAE_vs_Evidence}.} \label{fig2_RAE_vs_Evidence}
\end{figure}

From  Fig.~\ref{fig2_RAE_vs_Evidence}, there are clear correlations between ${\mathcal{S}_1}$ and ${\mathcal{I}}$ for all disciplines studied.
The corresponding values of Pearson coefficient are given in Table~\ref{tab_coefficients_2}.
The values of the correlation coefficients $r$ for the six disciplines studied here vary  from  0.87 to 0.96.
For comparison, the equivalent statistic for the biology  research groups studied in  Ref.~\cite{Evidence_2010} was $0.97$).
As in biology, the replacement of specific measures of quality and impact by their absolute counterparts has the effect of stretching  the corresponding axes by amounts proportional to the quantity of the groups or departments, and this leads to improved correlations.
 %
\begin{table}[!h]
\caption{The approximate values of the linear correlation coefficients between $\mathcal{S}_1$ and $\mathcal{I}$  for  several disciplines. Statistically significant values (with significance level $\alpha=0.05$) are highlighted in boldface.}
\begin{center}
\begin{tabular}{|l|l|l|l|}
\hline
Description of the  & \multicolumn{3}{|c|}{Pearson coefficient $r$}\\
\cline{2-4}
data sets &all &large   &medium\\
&groups&  groups&/small \\
&& &  groups\\
\hline
\hline
biology $^{\dag}$&
{\mathversion{bold}$0.97$}& 
{\mathversion{bold}$0.96$}& 
{\mathversion{bold}$0.90$}
\\
(44 groups: 32 large, &&&\\
7 medium, 5 small) &&&\\
\hline
chemistry&
{\mathversion{bold}$0.96$}& 
{\mathversion{bold}$0.96$}& 
 {\mathversion{bold}$0.79$}
\\
(29 groups: 12 large, &&&\\
14 medium, 3 small) &&&\\
\hline
physics&
{\mathversion{bold}$0.96$}& 
{\mathversion{bold}$0.96$}& 
{\mathversion{bold}$0.67$}
\\
(41 groups: 28 large, &&&\\
9 medium, 4 small)&&&\\
\hline
mechanical, aeronautical&
{\mathversion{bold}$0.92$}& 
--& 
--
\\
and manufacturing engineering&& &\\
(30 groups)&&&\\
\hline
geography and environmental &
{\mathversion{bold}$0.92$}& 
{\mathversion{bold}$0.56$}& 
{\mathversion{bold}$0.93$}
\\
studies &&&\\
(41 groups: 28 large, &&&\\
9 medium, 4 small)& &&\\
\hline
sociology &
{\mathversion{bold}$0.88$}& 
{\mathversion{bold}$0.82$}& 
{\mathversion{bold}$0.73$}
\\
(39 groups: 29 large, &&&\\
8 medium, 2 small) &&&\\
\hline
history &
{\mathversion{bold}$0.88$}& 
{\mathversion{bold}$0.79$}& 
{\mathversion{bold}$0.66$}
\\
(79 groups: 30 large, &&&\\
24 medium, 25 small) &&&\\
 \hline
\end{tabular}
\label{tab_coefficients_2}
\end{center}
\footnotesize{$^{\dag}$ The correlation coefficients for biology given in Ref.~\cite{2012_Scientometrics} were based on the overall
quality profiles $\mathcal{S}$. Here, to properly compare with the other subject areas and with $\mathcal{I}$,
the output-based absolute scores $\mathcal{S}_1$ are used instead.
}
\end{table}

As observed previously for biology \cite{2012_Scientometrics},
the correlation between $\mathcal{S}_1$ and $\mathcal{I}$ is usually best for large groups.
The only exception is geography: in this case medium and small groups exhibit a better correlation than large ones.
One may speculate as to the reasons for this. One possibility is the highly interdisciplinary nature of the research, which includes ``a wide range of enquiries into natural, environmental and human phenomena'' \cite{RAE_web}.
Indeed, among the disciplines analysed in this paper, only the geographical unit of assessment was declared as highly interdisciplinary and this marks it out.

While the $r$-values are high for all the disciplines, there is a noticeable difference between the  hard sciences (chemistry, physics and biology \cite{2012_Scientometrics})
and ``softer'' disciplined  (history and sociology).
For the former set, the correlation coefficient between absolute measures exceeds 95\%. For the latter set of disciplines it is smaller than 90\%.
The interdisciplinary area of geography and environmental studies with $r\approx 0.92$ is positioned somewhere between these two categories as is the engineering discipline studied.

\section{Conclusions}

Based on the above results, the following three main conclusions may be drawn.
\begin{itemize}
\item \emph{Weak correlations between specific measures of research quality and impact} have been observed for the disciplines of chemistry; physics; mechanical, aeronautical and manufacturing engineering; geography and environmental studies; sociology and history.
This signals that this citation-based measure is a {poor} proxy for peer-reviewed measures of the quality of research groups.
Moreover, since rankings are based on normalized data, this indicates that citation-based indicators will provide quite different rankings to those based on  peer review.
\item \emph{Strong correlation between absolute measures of research quality and impact} which was previously observed for biology  \cite{2012_Scientometrics}, is seen to extend to various extents to the disciplines which were analysed here.
Thus, citation-based measures may inform or serve as a  proxy for peer-review measures of the strengths of research groups.
\item Although the citation-based measures could be a reasonable  proxy {for}, or may inform about, the strengths of research groups for all  disciplines studied,
the results for the hard sciences are superior than those of the softer disciplines.
Specifically, Pearson coefficients exceeding 95\% were observed for physics, chemistry as well as for biology while the corresponding values for history and sociology are below 90\%. The interdisciplinary areas of geography and engineering are in between with a linear correlation of $\approx 92\%$ between absolute measures of scientific excellence.
\end{itemize}

Since quality-related funding is strength based, the use of citation-based indicators may offer a much cheaper, and less intrusive  alternative to the system currently in use in the UK and some other countries for large research groups in the hard sciences.
However, such a proxy would be far less reliable for the social sciences and humanities.
Moreover, citation-based indicators should not be used {in isolation} to compare the average quality of {Higher Education Institutions} or separate research groups. Nor should they be used for rankings.

\section*{Acknowledgements}
This work was supported in part by the 7th FP, IRSES project No. 269139 ``Dynamics and cooperative phenomena in complex physical and biological
environments'' and IRSES project No. 295302 ``Statistical physics in diverse realizations''.
The authors thank Jonathan Adams from {\emph{Thomson Reuters Research Analytics}} for the data and Ihor Mryglod for fruitful discussions.

\end{document}